\documentclass[prl, twocolumn, amsmath,floatfix, unsortedaddress, superscriptaddress]{revtex4-1}

\usepackage{graphicx}
\usepackage{epstopdf}
\usepackage{natbib}
\usepackage{bm}

\newcommand{\mfp}{{m\!f\:\!\!p}}

\newcommand{\PPPL}{Princeton Plasma Physics Laboratory, Princeton, NJ 08543, USA}
\newcommand{\Princeton}{Department of Astrophysical Sciences, Princeton University, Princeton, New Jersey 08540, USA}
\newcommand{\UofM}{Center for Ultrafast Optical Science, University of Michigan, Ann Arbor, Michigan 48109, USA}
\newcommand{\UofNH}{Space Science Center, University of New Hampshire, Durham, New Hampshire 03824, USA}
\newcommand{\UoF}{ Laboratoire de Physique des Plasmas, \'Ecole Polytechnique, Paris 75252, France}
\newcommand{\Laser}{Laboratory for Laser Energetics, University of Rochester, Rochester, New York 14623, USA}

\begin{document}

\begin{abstract}
Recent experiments have demonstrated magnetic reconnection between colliding plasma plumes, where the reconnecting magnetic fields were self-generated in the plasma by the Biermann battery effect. Using fully kinetic 3-D simulations, we show the full evolution of the magnetic fields and plasma in these experiments including self-consistent magnetic field generation about the expanding plume. The collision of the two plasmas drives the formation of a current sheet, where reconnection occurs in a strongly time-and space-dependent manner, demonstrating a new 3-D reconnection mechanism. Specifically, we observe fast, vertically-localized Biermann-mediated reconnection, an inherently 3-D process where the temperature profile in the current sheet coupled with the out-of-plane ablation density profile conspires to break inflowing field lines, reconnecting the field downstream. Fast reconnection is sustained by both the Biermann effect and the traceless electron pressure tensor, where the development of plasmoids appears to modulate the contribution of the latter. We present a simple and general formulation to consider the relevance of Biermann-mediated reconnection in general astrophysical scenarios. 

%\cite{NilsonPRL2006, LiPRL2007, WillingalePoP2010, ZhongNature2010, FikselPRL2014, RosenbergPRL2015}

\end{abstract}

% citations   \cite{Recon2D}  \cite{Biermann2D}  \cite{Vulcan}  \cite{SGII} 

%\pacs{}% insert suggested PACS numbers in braces on next line

\title{Biermann battery-mediated magnetic reconnection in 3-D colliding plasmas} 

\author{J. Matteucci}
\email{jmatteuc@pppl.gov}
\affiliation{\Princeton}
\author{W. Fox}
\affiliation{\Princeton}
\affiliation{\PPPL}
\author{A. Bhattacharjee}
\affiliation{\Princeton}
\affiliation{\PPPL}
\author{D. B. Schaeffer}
\affiliation{\Princeton}
\author{C. Moissard}
\affiliation{\UoF}
\author{K. Germaschewski}
\affiliation{\UofNH}
\author{G. Fiksel}
\affiliation{\UofM}
\author{S. X. Hu}
\affiliation{\Laser}
\date{\today}

\maketitle

The Biermann battery effect \cite{Biermann1950, KulsrudAPJ1997} is one of the few mechanisms known to spontaneously generate magnetic fields in plasmas. In the context of astrophysics, while too weak to generate the present-day observed cosmic magnetic fields, the Biermann battery effect is widely regarded as a possible source of the seed magnetic field, subsequently amplified to observed present-day fields by protogalactic turbulence \cite{KulsrudAPJ1997}. This effect, which is the result of non-collinearity between density and temperature gradients ($\frac{\partial \bm{B}}{\partial t} \propto \nabla n \times \nabla T$), has been shown in High Energy Density (HED) plasmas to generate strong magnetic fields (10-100 T) during intense laser heating \cite{NilsonPRL2006, LiPRL2007, WillingalePoP2010, ZhongNature2010, DongPRL2012, ManuelPRL2012, GaoPRL2012, RosenbergPRL2015}. In such experiments \cite{NilsonPRL2006, LiPRL2007, WillingalePoP2010, RosenbergPRL2015}, when two plumes are ablated adjacently, the oppositely polarized fields collide and undergo magnetic reconnection; this is the universal process in which the magnetic field threaded through plasma undergoes a fundamental topological change, often resulting in violent conversions of field energy into kinetic energy. HED laser plasmas provide a platform to study magnetic reconnection, which is intrinsic to many phenomena throughout plasma physics \cite{YamadaMPR2010}, from the sawtooth instability in magnetic fusion confinement devices \cite{Hastie1997} to solar flares \cite{LinSolPhys1976} and disturbances in Earth's magnetosphere \cite{PhanGRL2013}. 

Recent HED experiments have observed interesting effects associated with reconnection, including flux annihilation \cite{LiPRL2007, FikselPRL2014}, stagnation of reconnection \cite{RosenbergPRL2015}, and particle jets \cite{NilsonPRL2006, DongPRL2012}. Kinetic simulations of HED laser experiments based on model profiles have provided valuable insight into reconnection dynamics in this regime, including flux-pileup near the reconnection layer \cite{FoxPRL2011}, plasmoid formation \cite{FoxPoP2012b}, the role of the Nernst effect \cite{JoglekarPRL2014}, and particle acceleration by reconnection \cite{TotoricaPRL2016}. Particle-in-cell (PIC) simulations have also modeled Biermann-battery generation in expanding plasmas \cite{SchoefflerPRL2014}. One such study investigated reconnection in electron-dominated relativistic plasmas driven by short-pulse lasers \cite{PingPRE2014}, but to date the full 3-D evolution of magnetic reconnection within HED plasmas in the magnetohydrodynamic (MHD) regime (L / $d_{i}$ $\gg$ 1, where L is the system size and $d_{i}$ is the ion skin depth) has not been investigated. 

%This line of inquiry is important to understanding the possible role of magnetic fields and reconnection in indirect and direct-drive inertial confinement fusion experiments, as well as the physics of highly 3-dimensional reconnection systems in general \cite{RyggScience2008, LiScience2010}.
%, AndersonJGR1997

%This line of inquiry is important to understanding the possible role of magnetic fields and reconnection in indirect and direct-drive inertial confinement fusion experiments, as well as the physics of highly 3-dimensional reconnection systems in general \cite{RyggScience2008, LiScience2010}.

In this Letter, we present the first 3-D, end-to-end, fully kinetic computational study of magnetic reconnection within a recent HED experiment \cite{ZhongNature2010}, which captures both the \textit{self-consistent} initial field generation by the Biermann effect and reconnection within the fully 3-D geometry of the system. This is in contrast to previous HED reconnection simulations where the magnetic fields and plasma profiles are set in the initial conditions. We present in detail how reconnection proceeds, including how the development and ejection of plasmoids modulates the reconnection rate. The simulations also reveal that the Biermann battery effect can play a direct and significant role in 3-D magnetic reconnection, where a local $T_e$ maximum in the current sheet coupled with an out-of-plane density gradient conspires to reconnect flux via $\nabla n \times \nabla T$, a process we refer to as ``Biermann-mediated reconnection." We emphasize that observing this mechanism requires a full 3-D simulation, and it is distinct from previously-documented 2-D reconnection mechanisms, both in the strong guide-field case involving the electron scalar pressure \cite{WangJGR2000, FoxPRL2017} and in reconnection without a guide field involving the off-diagonal components of the electron pressure tensor \cite{HesseJGP1998, FoxPRL2011}. We introduce a new dimensionless number that enables the evaluation of the importance of the Biermann effect in reconnection and demonstrate its application to laboratory and space plasmas. 

For direct comparison, presented simulations are modeled after a recent reconnection experiment at the Shenguang-II (SG-II) laser facilities \cite{ZhongNature2010}, for which we observe general agreement between the field and plasma evolution in simulation and experiment. The results of our simulation have potentially important implications for reconnection and energy conversion in many 3-D reconnection systems. For example, in galactic dynamo theories where Biermann fields provide the seed magnetic field, it is implicitly assumed that fast reconnection occurs somehow to facilitate the breaking and reconfiguration needed to form large-scale magnetic fields \cite{KulsrudAPJ1997}. Our work demonstrates that in addition to seed generation, the Biermann effect might also facilitate the reconnection required for the success of a large-scale dynamo. Furthermore, this work is especially relevant to systems with large density and temperature gradients, such as in indirect and direct-drive inertial confinement fusion experiments \cite{RyggScience2008, LiScience2010} and reconnection in many astrophysical scenarios including in Earth's magnetosheath \cite{RetinoNP2007}, the heliopause \cite{OpherAPJ2011}, and in simulations of turbulent reconnection, including the highly turbulent reconnection upstream of high-Mach number shocks \cite{MatsumotoScience2015}.

We use the following formulation of generalized Ohm's law in order to quantitatively account for the contributions to the electric field, which in turn accounts for B-field evolution via Faraday's law,
%,  involved in the generation, advection, and reconnection of the magnetic field. Via Faraday's law and generalized Ohm's law, the evolution of the magnetic field is described by  
%\begin{equation}
%\label{eq:curlohm}
%\frac{\partial \bm{B}}{\partial t} \bm{=}   \nabla \times  \left[ (\bm{v} \times  \bm{B}) -\frac{\bm{j} \times  \bm{B}}{n_ee} + \frac{\nabla p_e}{n_ee} + \frac{\nabla \cdot \bm{\Pi_e}}{n_ee} + \frac{\bm{R}_{ei}}{n_ee} \right] ,
%\end{equation}

\begin{equation}
\label{eq:ohm}
\bm{E} \bm{=} - \bm{v} \times  \bm{B} +\frac{\bm{j} \times  \bm{B}}{n_ee} - \frac{\nabla p_e}{n_ee} - \frac{\nabla \cdot \bm{\Pi_e}}{n_ee} + \frac{\bm{R}_{ei}}{n_ee}.
\end{equation}
Each term on the RHS has a physical interpretation; the first two terms are the contribution due to ion flow and the Hall effect. The third term represents the contribution due to the scalar electron pressure. Via Faraday's law, this term contributes to $\frac{\partial \bm{B}}{\partial t}$ as the Biermann battery effect, $ - \frac{1}{n_ee} \nabla n_e \times \nabla T_e$, and is capable of generating or destroying magnetic flux depending on temperature, density, and field configurations. The fourth term, the traceless pressure tensor contribution (where $\bm{\Pi_e} \equiv \bm{P_e} - p_e \bm{I_3}$) is well-documented to be important in collisionless 2-D reconnection layers \cite{HesseJGP1998}. This term includes the effect of viscosity and contributions from pressure tensor anisotropy, which is associated with the Weibel instability \cite{WeibelPRL1958}. $\bm{R}_{ei}$ is the collisional momentum transfer between electrons and ions and via Faraday's law contributes to $\frac{\partial \bm{B}}{\partial t}$ as both resistive diffusion and Nernst advection via the thermal force. Noting that reconnection is characterized by a finite out-of-plane E-field inside the current sheet, we use Eq.\ \ref{eq:ohm} to quantify the contribution to the reconnection electric field from the Biermann-battery effect, which requires out-of-plane variation, and from the traceless pressure tensor term \cite{HesseJGP1998}.

%ion flow and Hall terms solely advect the field and therefore preserve magnetic field lines. The third term, which is the Biermann battery term, and which can be written as $ - \frac{1}{n_ee} \nabla n_e \times \nabla T_e$, generates (or destroys) field depending on the temperature, density, and field configurations. The fourth term, the traceless pressure tensor term (where $\bm{\Pi_e} \equiv \bm{P_e} - p_e \bm{I_3}$) is well-documented to be important in collisionless reconnection layers \cite{HesseJGP1998} and includes the effect of viscosity as well as contributions from anisotropies between on-diagonal pressure components, which are associated with the Weibel instability \cite{WeibelPRL1958}. $\bm{R}_{ei}$ is the collisional momentum transfer between electrons and ions and includes both resistive diffusion and advection via the Nernst effect (via the thermal force). Below, we use Eq.\ \ref{eq:curlohm} to characterize the role in the observed reconnection of the Biermann-battery effect, which depends solely on $T_e$ and $n_e$ profiles, versus the role of the traceless pressure tensor term, which requires off-diagonal components or on-diagonal anisotropy. Note, the latter term completely encapsulates the well-documented 2-D collisionless pressure tensor mechanisms \cite{HesseJGP1998}.

\begin{figure*}
\includegraphics[width= 18cm]{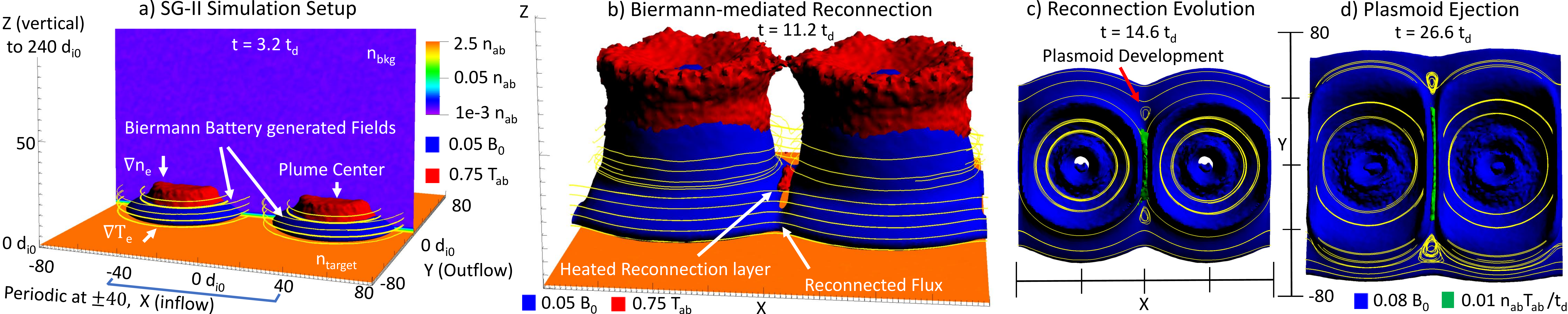}
\caption{Collision of self-magnetized plume with itself at four times, where two inflow ($X$) periods are shown. (a) Early time snapshot detailing setup of simulation; shown are two planar slices of $n_e$, one through the target ($z = 0 \,d_{i0}$) and one through the inflow, vertical plane (XZ, $y = 0 \,d_{i0}$), magnetic field lines in yellow, $|B|/B_0 = 0.05$ contour in blue, and $T_e/T_{ab} = 0.75$ contour in red. (b) Demonstration of Biermann-mediated reconnection at time t = $11.2 \,t_d$. (c, d) Top down views of later time field evolution (t = $14.6 \, t_d$ and $26.6 \, t_d$ respectively), including plasmoid ejection and current sheet stretching- here the blue contour represents $|B|/B_0 = 0.08$, and green represents $ \bm{E} \cdot  \bm{j} /(e n_{ab} T_{ab} /t_d) = 0.01$.}
\label{fig:Bigone}
\end{figure*}
%a model heating operator $\propto exp(\frac{x^2 + y^2}{R_h^2})$,
\textit{Simulation setup:} We use the PSC code \cite{GermaschewskiJCP2016}, a fully kinetic, explicit particle-in-cell code employing a binary Coulomb collision operator, to model the reference experiment. Shown in Fig.\ \ref{fig:Bigone}(a), simulations are initiated by heating electrons with a radial profile $H(x,y) \propto exp(\frac{-(x^2 +y^2)}{R_h^2})$ within a thin, dense target, producing an expanding plasma plume. All boundaries are periodic where the heated plume collides with itself along the inflow dimension ($X$), which is much shorter than the outflow ($Y$) and vertical ($Z$) dimensions. From the observed simulation ablation profiles we measure $n_{ab}$ and $T_{ab}$, the ablation density and temperature \cite{FoxArxiv2017}, where the heating operator magnitude is tuned to obtained the desired $T_{ab}$. We can then define the ablation ion skin depth, $d_{i0} = (M_i / n_{ab} Z e^2 \mu_0)^{1/2}$ and the sound speed, $C_{s, ab} = (Zk_bT_{ab}/M_i)^{1/2}$. Together these define the ablation timescale $t_d = d_{i0}/ C_{s, ab}$ and the characteristic magnetic field $B_0 = \sqrt{\mu_0n_{ab}k_b T_{ab}}$. Normalizing our simulations to ablation units allows us to match PSC results to physical values of $n_{ab, phys}$, $T_{ab, phys}$, $d_{i0, phys}$, and $t_{d, phys}$ obtained from a similar analysis of radiation-hydrodynamic simulations performed with the DRACO code \cite{SuxingPoP2013}. For reference, $n_{ab, phys}= 3\times 10^{27}$m$^{-3}$, $T_{ab, phys} = 2$ keV, $C_{ab, phys} \approx 300$ km/s, and $B_{0, phys} = 1300$ T. Ref.\ \cite{FoxArxiv2017} presents the SG-II experimental parameters and the full computational scheme for modeling heating, ablation, and matching to DRACO simulations. In contrast to Ref.\ \cite{FoxArxiv2017}, we extend our simulations from 2-D (XZ) to 3-D (XYZ), allowing reconnection.

We use a compressed electron-ion mass ratio, $Zm_e/M_i = 1/64$, ($Z=1$) and a compressed ratio between $T_{ab}$ and the electron rest mass energy $T_{ab}/m_ec^2 = 0.04$. We compress these values far below the physical ratios while achieving convergence in our results, provided $Zm_e/M_i, T_{ab}/m_ec^2 \ll 1$; in analogous 2-D simulations, ratios down to $Zm_e/M_i = 1/400$ show convergence with $Zm_e/M_i = 1/64$. A box of $80 \times 160 \times  480  \, d_{i0} (L_x, L_y, L_z)$ is used, with the grid cell spacing $\Delta x <  d_{e0}$, $5 \lambda_{D, ab}$, where $ \lambda_{D, ab} = \sqrt{\epsilon_0k_bT_{ab}/n_{ab}e^2}$. The SG-II heating radius $R_h/d_{i0} = 12$ with bubble separation $L_x/d_{i0} = 80$. Further parameters include the target width = $2\, d_{i0}$  the target density $=2.5 \, n_{ab}$, the background density $=0.001\, n_{ab}$, the target and background $T_e$ = 0.025 $T_{ab}$, and 50 second-order-shaped particles are used per cell at $n_{ab}$. The collisionality, described by $\lambda_{\mfp}$, the mean free path of electrons at $T_{ab}$ and $n_{ab}$, is matched to the electron skin depth $d_{e0}$ to perserve the correct collisional diffusivity of the magnetic field; $\lambda_{\mfp}/d_{e0} = 20$. While $\lambda_{\mfp}$ is resultingly mismatched on $d_{i0}$ scale, throughout the plume collision region we recover the correct collisionality regime, i.e. $\lambda_{mfp} > \delta_{rec}$, where $\delta_{rec}$ is the current sheet width. 

%The comparison beween the two setups shows how the evolution of reconnection depends on system size ($L/d_i$), where we expect more MHD-like behavior from the larger SG-II case.

%Applying Faraday's law to Eq 1. we find $> 95\%$ of the torodial magnetic field around the plume is generated by $ - \frac{1}{n_ee} \nabla n_e \times \nabla T_e$, the Biermann battery effect.

\begin{figure*}
\includegraphics[width= 18cm]{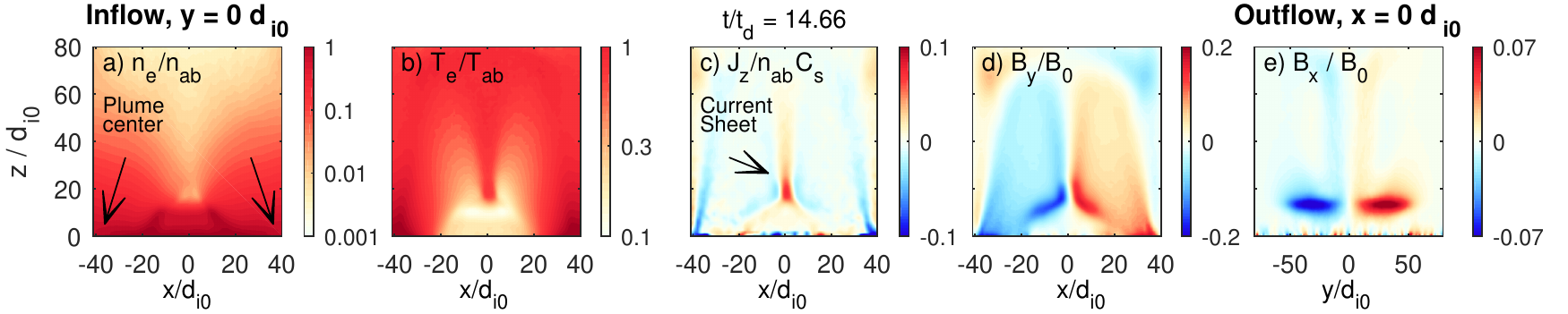}
\caption{ (a - d) 2-D slices along the inflow plane ($XZ$ plane, $ y = 0 \, d_{i0}$) of $n_e$, $T_e$, $J_z$, and $B_y$ at the SG-II scale at time $t = 14.6 \, t_d$, where the bottom corners correspond to the center of the expanding plume and the dotted black line indicates the current sheet. (e) 2-D slice along the outflow plane ($YZ$ plane, $x= 0 \, d_{i0}$) showing the downstream field $B_x$.}
\label{fig:longFig}
\end{figure*}

Fig.\ \ref{fig:Bigone} and Fig.\ \ref{fig:longFig} respectively show 3-D snapshots and 2-D profile slices of the SG-II magnetized plume collision. Fig.\ \ref{fig:Bigone} (a) demonstrates the plume ablation, where the magnetic field is generated via $ - \frac{1}{n_ee} \nabla n_e \times \nabla T_e$ due to the resulting density and temperature profiles, as demonstrated in Ref.\ \cite{FoxArxiv2017}. By $t = 11.2\, t_d$, seen in Fig.\ \ref{fig:Bigone} (b), the torodial fields have collided and begun to reconnect. In Fig.\ \ref{fig:Bigone} (c), at $t = 14.6\, t_d$, we observe field energy conversion in the reconnection current sheet between the inflow fields ($\bm{E} \cdot  \bm{j} > 0.01 \,n_{ab}T_{ab}/t_d$), and the development of closed-flux-surface, plasmoid-like structures in the outflow. The associated inflow plane profiles at $t = 14.6\, t_d$ of $n_e$, $T_e$, $J_z$, and inflow field $B_y$ are shown in Fig.\ \ref{fig:longFig} (a-d). From Fig.\ \ref{fig:longFig} (c,d), we find the upstream field at the edge of the current sheet is compressed to 0.17 $B_0$, a factor of 1.5-2$ \times$ the nominal generated field of $\approx 0.1 \,B_0$. This observation is in contrast to 2-D HED driven reconnection simulations, which find pile-up ratios of 4$\times$ the nominal inflow field \cite{FoxPRL2011}. In physical units, the observed compressed field corresponds to 220 T, comparable to the 370 T estimated in the experiment \cite{DongPRL2012}.

Fig.\ \ref{fig:longFig} (e) presents the downstream magnetic field in the outflow (YZ plane, $x = 0 \, d_{i0}$) at t = 14.6 $t_d$. For $z$ = 25-70 $d_{i0}$, moving in the $+y$ direction from the current sheet center at $y$ = 0 $d_{i0}$, we find $B_x$ reverses its sign twice, around $y$ = 25 $d_{i0}$ and  $y$ = 45 $d_{i0}$, where the first reversal corresponds to the center of the upper plasmoid in Fig.\ \ref{fig:Bigone} (c). Comparing Fig.\ \ref{fig:Bigone} (c) and (d), we find that the current sheet elongates as the plasmoids are ejected. Each plasmoid travels $\sim$ 20 $d_{i0}$ in 12 $t_d$, yielding an outflow speed of 1.66 $C_s$ $\approx 500$ km/s, in agreement with the experiment, which observed plasmoids ejected in the outflow at $400 \pm 50$ km/s \cite{ZhongNature2010}. Discussed below, the creation and ejection of plasmoids appears to modulate the reconnection, particularly via the traceless pressure tensor. Biermann-mediated reconnection appears to be independent of plasmoid behavior.

%, including a localized structure of flux at $ z \leq 10 \, d_{i0}$ and a more diffuse, extended structure from $z = 20 - 80  \, d_{i0}$. These two regions are clues to the multiple reconnection mechanisms operating in the simulation.

% showing the spatially dependent structure of the reconnected flux. 

\textit{Biermann-mediated Reconnection:} The profiles shown in Fig.\ \ref{fig:Bigone} (b) and Fig.\ \ref{fig:longFig} (a-d) demonstrate how Biermann-mediated reconnection operates. The current sheet (around $x \approx  0  \, d_{i0}, z \approx  20  \, d_{i0}$), is relatively heated compared to both the inflow and outflow, as shown in Fig.\ \ref{fig:Bigone} and \ref{fig:longFig} (b). This local $T_e$ maximum flips the direction of $\nabla T_e$ to point towards the plume collision center throughout the current sheet. Given $\nabla n$ remains directed toward the high-density target, $ - \frac{1}{n_ee} \nabla n_e \times \nabla T_e$ in the current sheet destroys incoming flux (XZ plane) and generates reconnected flux in the outflow (YZ plane).

%$ \bm{v} \times  \bm{B}$, $-\frac{\bm{j} \times  \bm{B}}{n_ee}$ $ \frac{\nabla p_e}{n_ee}$ $ \frac{\nabla \cdot \bm{\Pi_e}}{n_ee}$
\begin{figure}
\includegraphics[width= 9cm]{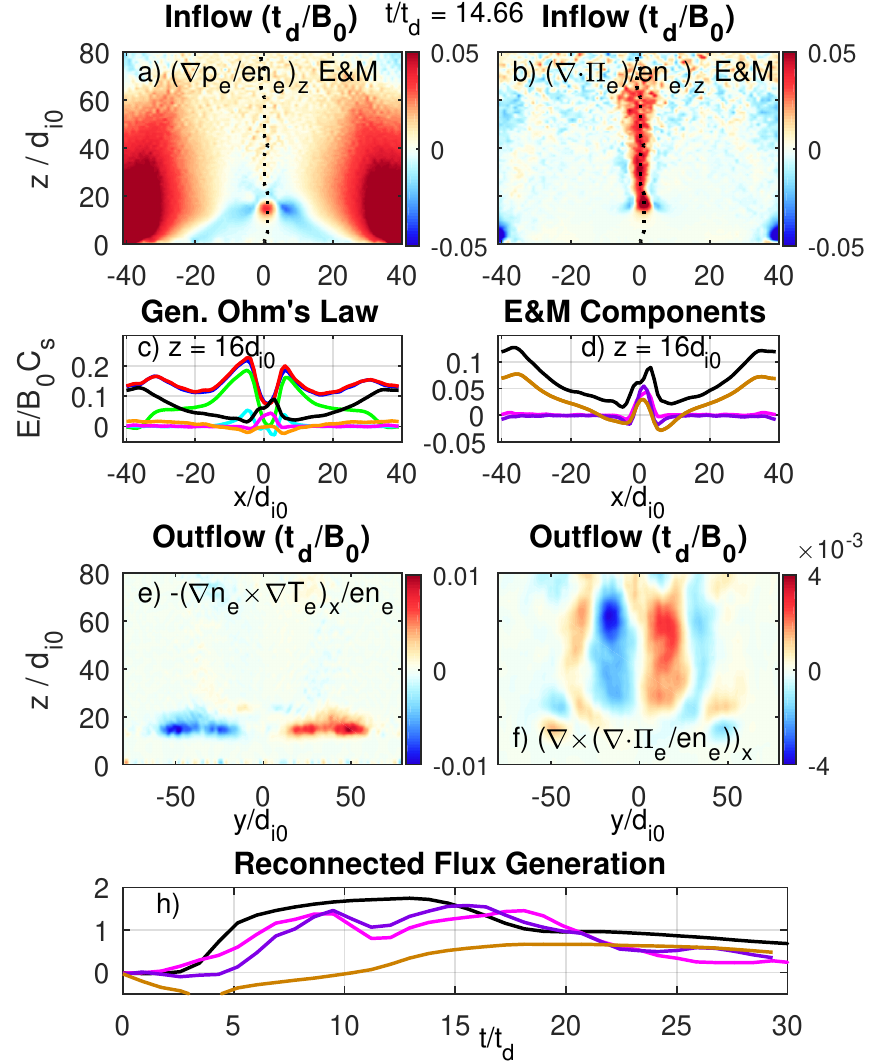}
\caption{(a,b) 2-D slice along the inflow plane of  $(\nabla p_e/en_e)_{z,em}$ and $(\nabla\cdot \Pi_e/en_e)_{z,em}$, respectively; dotted black line indicates the current sheet, ($B_y/B_0 = 0$). (c) Full out-of-plane generalized Ohm's law along the inflow at $y= 0 \,d_{i0}, z= 16 \,d_{i0}$: $E_z$ in blue, sum of RHS of Eq \ref{eq:ohm} in red, $\bm{v} \times  \bm{B}$ in green, $-\bm{j} \times  \bm{B}/n_ee$ in cyan, $\nabla p_e/en_e$ in black, $\nabla\cdot \Pi_e/en_e$ in magenta, and $R_{ei}/en_e$ in orange. (d) shows same cut of full z-component of  $\nabla p_e/en_e$ and $\nabla\cdot \Pi_e/en_e$ (black and magenta, respectively) vs. EM component of each term (brown and purple, respectively.) (e,f) show the downstream flux creation in the outflow (YZ, $x = 0 \,d_{i0}$) by  $ - \frac{1}{n_ee} \nabla n_e \times \nabla T_e$ and  $ -\nabla \times (\nabla \cdot \Pi_e/en_e)$, respectively. The quantities in (a-f) are calculated based on 2 $d_{i0}$ averaging of plasma parameters in all 3 dimensions. (g) shows the rate of flux reconnection, $\dot{\Psi}= \int_{C}E_{z,em} dz$, where $C$ is along the current sheet, by $(\nabla p_e/en_e)_{z,em}$ (brown) and $(\nabla\cdot \Pi_e/en_e)_{z,em}$ (purple.) (g,h) also show $\dot{\Psi}= \int_{0}^{\frac{L_z}{2}}\int_{0}^{\frac{L_y}{2}} \frac{\partial B_{x}}{\partial t} dydz$, the rate of flux creation in the outflow by Biermann, $ - \frac{1}{n_ee} \nabla n_e \times \nabla T_e$, (black) and the traceless pressure tensor term $ -\nabla \times (\nabla \cdot \Pi_e/en_e)$ (magenta.) }
\label{fig:dbdt}
\end{figure}

In Fig.\ \ref{fig:dbdt}, we quantify the reconnection contribution of both the Biermann and traceless pressure tensor mechanisms. In 3-D reconnection, the out-of-plane electric field $E_z$ can have an electrostatic contribution in the current sheet which does not contribute to B-field evolution (i.e. reconnection.) Therefore, we extract the electromagnetic (EM) contribution of each term in Eq.\ 1 by solving for the divergence-less component of each RHS term via Fourier analysis. In Fig.\ \ref{fig:dbdt} (a,b), the electromagnetic contributions to $E_{z,em}$ of the two mechanisms along the inflow (XZ) plane are presented, demonstrating $(\nabla p_e/en_e)_{em}$ contributes significantly to the reconnection E-field in the current sheet (black dotted line) around $z \approx 16\,d_{i0}$, while $(\nabla\cdot \Pi_e/en_e)_{em}$ operates throughout the vertical current sheet. Fig.\ \ref{fig:dbdt} (c) presents the full generalized Ohm's law for $E_z$ at $z = 16\,d_{i0}$, showing strong ion inflow (green), Hall inflow (cyan), dissipation at $x= 0 \, d_{i0}$ in the current sheet due to both $\nabla p_e/en_e$ (black) and $\nabla\cdot \Pi_e/en_e$ (magenta), and agreement between $E_z$ and the sum of the RHS of Eq.\ \ref{eq:ohm} (blue and red, respectively.) Fig.\ \ref{fig:dbdt} (d) clarifies the dissipation terms, presenting both the full and EM contributions to Eq.\ 1. Comparing $\nabla p_e/en_e$ (black) vs. $(\nabla p_e/en_e)_{z, em}$ (brown), we find the scalar pressure term has both significant electromagnetic and electrostatic components in the current sheet with $(\nabla p_e/en_e)_{z, em}$ = 0.03 $B_0C_s$. For the traceless pressure tensor term, we find $\nabla\cdot \Pi_e/en_e \approx (\nabla\cdot \Pi_e/en_e)_{z, em} =$ 0.055 $B_0C_s$. We define the local Alfvenic reconnection rate normalization as $B_{up}^* V_{A,up}^*$, where the maximum upstream field $B_{up}^*$ = 0.17 $B_0$ and the Alfven velocity $V_A^*$ = $B_{up}^*/\sqrt{\mu_0 M_i n_i^*} = 0.85 \, C_s$, where $n_i^* = 0.04 \, n_{ab}$ is the current sheet ion density at $z= 16 \, d_{i0}$. Noting $B_{up}^* V_{A,up}^* = 0.145 \, B_0 C_s$, we find the reconnection rate due to Biermann, $R_{Biermann} =  (\nabla p_e/en_e)_{z, em}/B_{up}^* V_{A,up}^* \approx 0.2$ and $R_{traceless} =  (\nabla\cdot \Pi_e/en_e)_{z, em}/B_{up}^* V_{A,up}^* \approx 0.38$. The local reconnection rate is the sum of these rates, yielding $R \approx 0.58$, which is the maximum rate observed in the simulation. 

Fig.\ \ref{fig:dbdt} (e,f) shows the outflow (YZ plane) B-field generation ($\frac{\partial B_x }{\partial t}$) due to $ - \frac{1}{n_ee} \nabla n_e \times \nabla T_e$, Biermann, and  $ -\nabla \times (\nabla \cdot \Pi_e/en_e)$, the traceless pressure tensor, where the latter appears to be associated with plasmoid formation from $z$ = 25-80 $d_{i0}$. Fig.\ \ref{fig:dbdt} (g,f) quantifies the full picture: the rate of total flux reconnection, calculated by $(\nabla p_e/en_e)_{em}$ and $(\nabla\cdot \Pi_e/en_e)_{em}$ vertically integrated along the current sheet (brown, purple respectively) vs.\ time, as well as the downstream flux generation rate by Biermann, as in Fig.\ \ref{fig:dbdt} (e), and the traceless pressure tensor, as in Fig.\ \ref{fig:dbdt} (f), integrated over the outflow (respectively shown in black and magenta.) For the traceless pressure tensor, the reconnected flux (purple) matches the flux creation downstream (magenta) at both scales, and we find global reconnection to speed up around t = $14 \, t_d$, corresponding to the plasmoid creation. Subsequently (t = $17-30 \, t_d$), reconnection progressively slows as the plasmoids are ejected and the current sheet lengthens. For the Biermann term, we find reconnection (brown) does not match downstream flux generation (black), indicating that some downstream flux generation by the Biermann effect occurs separately from the reconnection. Biermann reconnection does not slow with current sheet elongation and ultimately constitutes $\approx$ 25\% of the total reconnection, whereas the Biermann term generates roughly 50\% of the downstream flux.

\textit{Biermann Reconnection Rate:}
To evaluate when the Biermann battery effect should be considered in a reconnection scenario, we compare estimates of $R_{Biermann}$ against the typical fast reconnection value of  $0.1 \, B_{up} V_{A}$ \cite{ComissoJPP2016}. The Biermann mechanism requires a heated reconnection layer $T_e$ with an associated inflow scale length $L_T$, coupled with a significant out-of-plane $n_e$ variation with scale length $L_n$. Considering $ - \frac{1}{n_ee} \nabla n_e \times \nabla T_e$ operating over a reconnection layer width of $\delta_{rec}$, the rate of flux destruction in the reconnection layer, normalized to $B_{up}^* V_A^*$, can be approximated as:
\begin{equation}
\label{eq:Rb1}
R_{Biermann} \approx  \frac{\delta_{rec}T_e }{eL_T L_n B_{up}^* V_{A}^*} = \frac{\beta_e}{2} \, \frac{\delta_{rec} \, d_{i,rec}} {L_T \, L_n},
\end{equation}
where in the second equality we have substituted $\beta_e = n_e T_e /\frac{B_{up}^{*2}}{2 \mu_0}$, and $d_{i, rec}$ as the local ion inertial length, demonstrating $R_{Biermann}$ scales with reconnection layer $\beta_e$. In our simulations for $t = 14 \, t_d$, $z =16\,  d_{i0}$, given $T_{e, rec} \approx T_{ab}$, $\beta_e \approx 2.8$,  $L_T \approx 10 \, d_{i0}$, $\delta_{rec} \approx d_{i, rec} = 5 \, d_{i0}$, $L_n \approx 14 \, d_{i0}$, we find $R_{Biermann} \approx 0.25$, in agreement with results above.

From Ref.\ \cite{RetinoNP2007}, we use $R_{Biermann}$ to evaluate the possibility of Biermann-mediated reconnection within the turbulent magnetosheath. Referring to the current sheet parameters presented therein, we find $B_{in} \approx$ 20 nT, $V_A \approx 200$ km/s, $T_{e, rec} \approx$ 100 eV, $L_T \approx$ 180 km, $\delta_{rec} \approx$ 90 km, and we estimate $L_n \approx$ 240 km given associated density measurements. These estimates yield $R_{Biermann} \approx$ 0.05, comparable to the observed normalized reconnection rate of 0.1, demonstrating that Biermann reconnection may play a significant role in this system \cite{RetinoNP2007}.

%The simulations in this Letter present novel insights into the 3-D physics of laser-driven reconnection experiments. In addition to MHD-scale HED experiments, we propose that Biermann-mediated reconnection is likely important in recent short-pulse experiments, where the relativistic electrons are constrained to travel along the target surface yielding large out-of-plane pressure gradients which suggests a strong Biermann effect during reconnection \cite{RaymondArXiv2016}. Furthermore, Biermann-mediated reconnection may play a role in turbulent 3-D reconnection scenarios such as in the magnetosheath, which may be observed by the Magnetospheric Multiscale mission, and in the heliosheath.

The simulations presented in this Letter go beyond previous studies by capturing the full end-to-end 3-D evolution, from self-consistent field generation to reconnection, of a recent MHD-scale HED reconnection experiment. We find general agreement in B-field generation and plasmoid dynamics between simulation and experiment, and detailed analysis of our results reveals that Biermann-battery effect plays a direct role in the 3-D fast magnetic reconnection. In addition to in HED regimes, we show using the dimensionless parameter $R_{Biermann}$ that Biermann-mediated reconnection may play a role in turbulent 3-D reconnection within the magnetosheath, which may be observed by the Magnetospheric Multiscale mission. We further propose that this effect could be important in heliosheath and in mediating the initial fast reconnection necessary for large-scale dynamo. 

\begin{acknowledgments}
Simulations were conducted on the Titan supercomputer at the Oak Ridge Leadership Computing Facility at the Oak Ridge National Laboratory, supported by the Office of Science of the DOE under Contract No. DE-AC05-00OR22725. J. M. was supported by the DOD through the NDSEG Program, 32 CFR 168a. This research was also supported by the DOE under Contracts No. DE-SC0006670, No. DE-SC0008655, and No. DE-SC0016249.
\end{acknowledgments}
\bibliographystyle{apsrev4-1}

\bibliography{bibone_short,citeulike_short}

\begin{thebibliography}{34}%
\makeatletter
\providecommand \@ifxundefined [1]{%
 \@ifx{#1\undefined}
}%
\providecommand \@ifnum [1]{%
 \ifnum #1\expandafter \@firstoftwo
 \else \expandafter \@secondoftwo
 \fi
}%
\providecommand \@ifx [1]{%
 \ifx #1\expandafter \@firstoftwo
 \else \expandafter \@secondoftwo
 \fi
}%
\providecommand \natexlab [1]{#1}%
\providecommand \enquote  [1]{``#1''}%
\providecommand \bibnamefont  [1]{#1}%
\providecommand \bibfnamefont [1]{#1}%
\providecommand \citenamefont [1]{#1}%
\providecommand \href@noop [0]{\@secondoftwo}%
\providecommand \href [0]{\begingroup \@sanitize@url \@href}%
\providecommand \@href[1]{\@@startlink{#1}\@@href}%
\providecommand \@@href[1]{\endgroup#1\@@endlink}%
\providecommand \@sanitize@url [0]{\catcode `\\12\catcode `\$12\catcode
  `\&12\catcode `\#12\catcode `\^12\catcode `\_12\catcode `\%12\relax}%
\providecommand \@@startlink[1]{}%
\providecommand \@@endlink[0]{}%
\providecommand \url  [0]{\begingroup\@sanitize@url \@url }%
\providecommand \@url [1]{\endgroup\@href {#1}{\urlprefix }}%
\providecommand \urlprefix  [0]{URL }%
\providecommand \Eprint [0]{\href }%
\providecommand \doibase [0]{http://dx.doi.org/}%
\providecommand \selectlanguage [0]{\@gobble}%
\providecommand \bibinfo  [0]{\@secondoftwo}%
\providecommand \bibfield  [0]{\@secondoftwo}%
\providecommand \translation [1]{[#1]}%
\providecommand \BibitemOpen [0]{}%
\providecommand \bibitemStop [0]{}%
\providecommand \bibitemNoStop [0]{.\EOS\space}%
\providecommand \EOS [0]{\spacefactor3000\relax}%
\providecommand \BibitemShut  [1]{\csname bibitem#1\endcsname}%
\let\auto@bib@innerbib\@empty
%</preamble>
\bibitem [{\citenamefont {Biermann}(1950)}]{Biermann1950}%
  \BibitemOpen
  \bibfield  {author} {\bibinfo {author} {\bibfnamefont {L.}~\bibnamefont
  {Biermann}},\ }\href@noop {} {\bibfield  {journal} {\bibinfo  {journal} {Z.
  Naturforsch}\ }\textbf {\bibinfo {volume} {5a}},\ \bibinfo {pages} {65}
  (\bibinfo {year} {1950})}\BibitemShut {NoStop}%
\bibitem [{\citenamefont {Kulsrud}\ \emph {et~al.}(1997)\citenamefont
  {Kulsrud}, \citenamefont {Cen}, \citenamefont {Ostriker},\ and\ \citenamefont
  {Ryutov}}]{KulsrudAPJ1997}%
  \BibitemOpen
  \bibfield  {author} {\bibinfo {author} {\bibfnamefont {R.~M.}\ \bibnamefont
  {Kulsrud}}, \bibinfo {author} {\bibfnamefont {R.}~\bibnamefont {Cen}},
  \bibinfo {author} {\bibfnamefont {J.~P.}\ \bibnamefont {Ostriker}}, \ and\
  \bibinfo {author} {\bibfnamefont {D.}~\bibnamefont {Ryutov}},\ }\href
  {http://stacks.iop.org/0004-637X/480/i=2/a=481} {\bibfield  {journal}
  {\bibinfo  {journal} {Astrophys. J.}\ }\textbf {\bibinfo {volume} {480}},\
  \bibinfo {pages} {481} (\bibinfo {year} {1997})}\BibitemShut {NoStop}%
\bibitem [{\citenamefont {Nilson}\ \emph {et~al.}(2006)\citenamefont {Nilson},
  \citenamefont {Willingale}, \citenamefont {Kaluza} \emph
  {et~al.}}]{NilsonPRL2006}%
  \BibitemOpen
  \bibfield  {author} {\bibinfo {author} {\bibfnamefont {P.~M.}\ \bibnamefont
  {Nilson}}, \bibinfo {author} {\bibfnamefont {L.}~\bibnamefont {Willingale}},
  \bibinfo {author} {\bibfnamefont {M.~C.}\ \bibnamefont {Kaluza}},  \emph
  {et~al.},\ }\href@noop {} {\bibfield  {journal} {\bibinfo  {journal} {Phys.
  Rev. Lett.}\ }\textbf {\bibinfo {volume} {97}},\ \bibinfo {pages} {255001}
  (\bibinfo {year} {2006})}\BibitemShut {NoStop}%
\bibitem [{\citenamefont {Li}\ \emph {et~al.}(2007)\citenamefont {Li},
  \citenamefont {S\'eguin}, \citenamefont {Frenje} \emph {et~al.}}]{LiPRL2007}%
  \BibitemOpen
  \bibfield  {author} {\bibinfo {author} {\bibfnamefont {C.~K.}\ \bibnamefont
  {Li}}, \bibinfo {author} {\bibfnamefont {F.~H.}\ \bibnamefont {S\'eguin}},
  \bibinfo {author} {\bibfnamefont {J.~A.}\ \bibnamefont {Frenje}},  \emph
  {et~al.},\ }\href@noop {} {\bibfield  {journal} {\bibinfo  {journal} {Phys.
  Rev. Lett.}\ }\textbf {\bibinfo {volume} {99}},\ \bibinfo {pages} {055001}
  (\bibinfo {year} {2007})}\BibitemShut {NoStop}%
\bibitem [{\citenamefont {Willingale}\ \emph {et~al.}(2010)\citenamefont
  {Willingale}, \citenamefont {Nilson}, \citenamefont {Kaluza} \emph
  {et~al.}}]{WillingalePoP2010}%
  \BibitemOpen
  \bibfield  {author} {\bibinfo {author} {\bibfnamefont {L.}~\bibnamefont
  {Willingale}}, \bibinfo {author} {\bibfnamefont {P.~M.}\ \bibnamefont
  {Nilson}}, \bibinfo {author} {\bibfnamefont {M.~C.}\ \bibnamefont {Kaluza}},
  \emph {et~al.},\ }\href {\doibase 10.1063/1.3377787} {\bibfield  {journal}
  {\bibinfo  {journal} {Phys. Plasmas}\ }\textbf {\bibinfo {volume} {17}},\
  \bibinfo {pages} {043104} (\bibinfo {year} {2010})}\BibitemShut {NoStop}%
\bibitem [{\citenamefont {Zhong}\ \emph {et~al.}(2010)\citenamefont {Zhong},
  \citenamefont {Li}, \citenamefont {Wang} \emph {et~al.}}]{ZhongNature2010}%
  \BibitemOpen
  \bibfield  {author} {\bibinfo {author} {\bibfnamefont {J.}~\bibnamefont
  {Zhong}}, \bibinfo {author} {\bibfnamefont {Y.}~\bibnamefont {Li}}, \bibinfo
  {author} {\bibfnamefont {X.}~\bibnamefont {Wang}},  \emph {et~al.},\ }\href
  {\doibase 10.1038/nphys1790} {\bibfield  {journal} {\bibinfo  {journal}
  {Nature Physics}\ }\textbf {\bibinfo {volume} {6}},\ \bibinfo {pages} {984}
  (\bibinfo {year} {2010})}\BibitemShut {NoStop}%
\bibitem [{\citenamefont {Dong}\ \emph {et~al.}(2012)\citenamefont {Dong},
  \citenamefont {Wang}, \citenamefont {Lu} \emph {et~al.}}]{DongPRL2012}%
  \BibitemOpen
  \bibfield  {author} {\bibinfo {author} {\bibfnamefont {Q.~L.}\ \bibnamefont
  {Dong}}, \bibinfo {author} {\bibfnamefont {S.~J.}\ \bibnamefont {Wang}},
  \bibinfo {author} {\bibfnamefont {Q.~M.}\ \bibnamefont {Lu}},  \emph
  {et~al.},\ }\href {\doibase 10.1103/physrevlett.108.215001} {\bibfield
  {journal} {\bibinfo  {journal} {Phys. Rev. Lett.}\ }\textbf {\bibinfo
  {volume} {108}},\ \bibinfo {pages} {215001} (\bibinfo {year}
  {2012})}\BibitemShut {NoStop}%
\bibitem [{\citenamefont {Manuel}\ \emph {et~al.}(2012)\citenamefont {Manuel},
  \citenamefont {Li}, \citenamefont {S\'eguin} \emph {et~al.}}]{ManuelPRL2012}%
  \BibitemOpen
  \bibfield  {author} {\bibinfo {author} {\bibfnamefont {M.~J.-E.}\
  \bibnamefont {Manuel}}, \bibinfo {author} {\bibfnamefont {C.~K.}\
  \bibnamefont {Li}}, \bibinfo {author} {\bibfnamefont {F.~H.}\ \bibnamefont
  {S\'eguin}},  \emph {et~al.},\ }\href@noop {} {\bibfield  {journal} {\bibinfo
   {journal} {Phys. Rev. Lett.}\ }\textbf {\bibinfo {volume} {108}},\ \bibinfo
  {pages} {255006} (\bibinfo {year} {2012})}\BibitemShut {NoStop}%
\bibitem [{\citenamefont {Gao}\ \emph {et~al.}(2012)\citenamefont {Gao},
  \citenamefont {Nilson}, \citenamefont {Igumenschev} \emph
  {et~al.}}]{GaoPRL2012}%
  \BibitemOpen
  \bibfield  {author} {\bibinfo {author} {\bibfnamefont {L.}~\bibnamefont
  {Gao}}, \bibinfo {author} {\bibfnamefont {P.~M.}\ \bibnamefont {Nilson}},
  \bibinfo {author} {\bibfnamefont {I.~V.}\ \bibnamefont {Igumenschev}},  \emph
  {et~al.},\ }\href@noop {} {\bibfield  {journal} {\bibinfo  {journal} {Phys.
  Rev. Lett.}\ }\textbf {\bibinfo {volume} {109}},\ \bibinfo {pages} {115001}
  (\bibinfo {year} {2012})}\BibitemShut {NoStop}%
\bibitem [{\citenamefont {Fiksel}\ \emph {et~al.}(2014)\citenamefont {Fiksel},
  \citenamefont {Fox}, \citenamefont {Bhattacharjee} \emph
  {et~al.}}]{FikselPRL2014}%
  \BibitemOpen
  \bibfield  {author} {\bibinfo {author} {\bibfnamefont {G.}~\bibnamefont
  {Fiksel}}, \bibinfo {author} {\bibfnamefont {W.}~\bibnamefont {Fox}},
  \bibinfo {author} {\bibfnamefont {A.}~\bibnamefont {Bhattacharjee}},  \emph
  {et~al.},\ }\href {\doibase 10.1103/PhysRevLett.113.105003} {\bibfield
  {journal} {\bibinfo  {journal} {Phys. Rev. Lett.}\ }\textbf {\bibinfo
  {volume} {113}},\ \bibinfo {pages} {105003} (\bibinfo {year}
  {2014})}\BibitemShut {NoStop}%
\bibitem [{\citenamefont {Rosenberg}\ \emph {et~al.}(2015)\citenamefont
  {Rosenberg}, \citenamefont {Li}, \citenamefont {Fox} \emph
  {et~al.}}]{RosenbergPRL2015}%
  \BibitemOpen
  \bibfield  {author} {\bibinfo {author} {\bibfnamefont {M.~J.}\ \bibnamefont
  {Rosenberg}}, \bibinfo {author} {\bibfnamefont {C.~K.}\ \bibnamefont {Li}},
  \bibinfo {author} {\bibfnamefont {W.}~\bibnamefont {Fox}},  \emph {et~al.},\
  }\href {\doibase 10.1103/physrevlett.114.205004} {\bibfield  {journal}
  {\bibinfo  {journal} {Phys. Rev. Lett.}\ }\textbf {\bibinfo {volume} {114}},\
  \bibinfo {pages} {205004} (\bibinfo {year} {2015})}\BibitemShut {NoStop}%
\bibitem [{\citenamefont {Yamada}\ \emph {et~al.}(2010)\citenamefont {Yamada},
  \citenamefont {Kulsrud},\ and\ \citenamefont {Ji}}]{YamadaMPR2010}%
  \BibitemOpen
  \bibfield  {author} {\bibinfo {author} {\bibfnamefont {M.}~\bibnamefont
  {Yamada}}, \bibinfo {author} {\bibfnamefont {R.}~\bibnamefont {Kulsrud}}, \
  and\ \bibinfo {author} {\bibfnamefont {H.}~\bibnamefont {Ji}},\ }\href
  {\doibase 10.1103/RevModPhys.82.603} {\bibfield  {journal} {\bibinfo
  {journal} {Rev. Mod. Phys.}\ }\textbf {\bibinfo {volume} {82}},\ \bibinfo
  {pages} {603} (\bibinfo {year} {2010})}\BibitemShut {NoStop}%
\bibitem [{\citenamefont {Hastie}(1997)}]{Hastie1997}%
  \BibitemOpen
  \bibfield  {author} {\bibinfo {author} {\bibfnamefont {R.~J.}\ \bibnamefont
  {Hastie}},\ }\href {\doibase 10.1023/A:1001728227899} {\bibfield  {journal}
  {\bibinfo  {journal} {Astrophys. and Space Sci.}\ }\textbf {\bibinfo {volume}
  {256}},\ \bibinfo {pages} {177} (\bibinfo {year} {1997})}\BibitemShut
  {NoStop}%
\bibitem [{\citenamefont {Lin}\ and\ \citenamefont
  {Hudson}(1976)}]{LinSolPhys1976}%
  \BibitemOpen
  \bibfield  {author} {\bibinfo {author} {\bibfnamefont {R.~P.}\ \bibnamefont
  {Lin}}\ and\ \bibinfo {author} {\bibfnamefont {H.~S.}\ \bibnamefont
  {Hudson}},\ }\bibfield  {booktitle} {\emph {\bibinfo {booktitle} {Solar
  Physics}},\ }\href {\doibase 10.1007/bf00206199} {\ \textbf {\bibinfo
  {volume} {50}},\ \bibinfo {pages} {153} (\bibinfo {year} {1976})}\BibitemShut
  {NoStop}%
\bibitem [{\citenamefont {Phan}\ \emph {et~al.}(2013)\citenamefont {Phan},
  \citenamefont {Shay}, \citenamefont {Gosling} \emph {et~al.}}]{PhanGRL2013}%
  \BibitemOpen
  \bibfield  {author} {\bibinfo {author} {\bibfnamefont {T.~D.}\ \bibnamefont
  {Phan}}, \bibinfo {author} {\bibfnamefont {M.~A.}\ \bibnamefont {Shay}},
  \bibinfo {author} {\bibfnamefont {J.~T.}\ \bibnamefont {Gosling}},  \emph
  {et~al.},\ }\href {\doibase 10.1002/grl.50917} {\bibfield  {journal}
  {\bibinfo  {journal} {Geophys. Res. Lett.}\ }\textbf {\bibinfo {volume}
  {40}},\ \bibinfo {pages} {4475} (\bibinfo {year} {2013})}\BibitemShut
  {NoStop}%
\bibitem [{\citenamefont {Fox}\ \emph {et~al.}(2011)\citenamefont {Fox},
  \citenamefont {Bhattacharjee},\ and\ \citenamefont
  {Germaschewski}}]{FoxPRL2011}%
  \BibitemOpen
  \bibfield  {author} {\bibinfo {author} {\bibfnamefont {W.}~\bibnamefont
  {Fox}}, \bibinfo {author} {\bibfnamefont {A.}~\bibnamefont {Bhattacharjee}},
  \ and\ \bibinfo {author} {\bibfnamefont {K.}~\bibnamefont {Germaschewski}},\
  }\href {\doibase 10.1103/PhysRevLett.106.215003} {\bibfield  {journal}
  {\bibinfo  {journal} {Phys. Rev. Lett.}\ }\textbf {\bibinfo {volume} {106}},\
  \bibinfo {pages} {215003} (\bibinfo {year} {2011})}\BibitemShut {NoStop}%
\bibitem [{\citenamefont {Fox}\ \emph {et~al.}(2012)\citenamefont {Fox},
  \citenamefont {Bhattacharjee},\ and\ \citenamefont
  {Germaschewski}}]{FoxPoP2012b}%
  \BibitemOpen
  \bibfield  {author} {\bibinfo {author} {\bibfnamefont {W.}~\bibnamefont
  {Fox}}, \bibinfo {author} {\bibfnamefont {A.}~\bibnamefont {Bhattacharjee}},
  \ and\ \bibinfo {author} {\bibfnamefont {K.}~\bibnamefont {Germaschewski}},\
  }\href {\doibase 10.1063/1.3694119} {\bibfield  {journal} {\bibinfo
  {journal} {Phys. Plasmas}\ }\textbf {\bibinfo {volume} {19}},\ \bibinfo
  {pages} {056309} (\bibinfo {year} {2012})}\BibitemShut {NoStop}%
\bibitem [{\citenamefont {Joglekar}\ \emph {et~al.}(2014)\citenamefont
  {Joglekar}, \citenamefont {Thomas}, \citenamefont {Fox},\ and\ \citenamefont
  {Bhattacharjee}}]{JoglekarPRL2014}%
  \BibitemOpen
  \bibfield  {author} {\bibinfo {author} {\bibfnamefont {A.~S.}\ \bibnamefont
  {Joglekar}}, \bibinfo {author} {\bibfnamefont {A.~G.~R.}\ \bibnamefont
  {Thomas}}, \bibinfo {author} {\bibfnamefont {W.}~\bibnamefont {Fox}}, \ and\
  \bibinfo {author} {\bibfnamefont {A.}~\bibnamefont {Bhattacharjee}},\ }\href
  {\doibase 10.1103/physrevlett.112.105004} {\bibfield  {journal} {\bibinfo
  {journal} {Phys. Rev. Lett.}\ }\textbf {\bibinfo {volume} {112}},\ \bibinfo
  {pages} {105004} (\bibinfo {year} {2014})}\BibitemShut {NoStop}%
\bibitem [{\citenamefont {Totorica}\ \emph {et~al.}(2016)\citenamefont
  {Totorica}, \citenamefont {Abel},\ and\ \citenamefont
  {Fiuza}}]{TotoricaPRL2016}%
  \BibitemOpen
  \bibfield  {author} {\bibinfo {author} {\bibfnamefont {S.~R.}\ \bibnamefont
  {Totorica}}, \bibinfo {author} {\bibfnamefont {T.}~\bibnamefont {Abel}}, \
  and\ \bibinfo {author} {\bibfnamefont {F.}~\bibnamefont {Fiuza}},\ }\href
  {\doibase 10.1103/physrevlett.116.095003} {\bibfield  {journal} {\bibinfo
  {journal} {Phys. Rev. Lett.}\ }\textbf {\bibinfo {volume} {116}},\ \bibinfo
  {pages} {095003} (\bibinfo {year} {2016})}\BibitemShut {NoStop}%
\bibitem [{\citenamefont {Schoeffler}\ \emph {et~al.}(2014)\citenamefont
  {Schoeffler}, \citenamefont {Loureiro}, \citenamefont {Fonseca},\ and\
  \citenamefont {Silva}}]{SchoefflerPRL2014}%
  \BibitemOpen
  \bibfield  {author} {\bibinfo {author} {\bibfnamefont {K.~M.}\ \bibnamefont
  {Schoeffler}}, \bibinfo {author} {\bibfnamefont {N.~F.}\ \bibnamefont
  {Loureiro}}, \bibinfo {author} {\bibfnamefont {R.~A.}\ \bibnamefont
  {Fonseca}}, \ and\ \bibinfo {author} {\bibfnamefont {L.~O.}\ \bibnamefont
  {Silva}},\ }\href {\doibase 10.1103/physrevlett.112.175001} {\bibfield
  {journal} {\bibinfo  {journal} {Phys. Rev. Lett.}\ }\textbf {\bibinfo
  {volume} {112}},\ \bibinfo {pages} {175001} (\bibinfo {year}
  {2014})}\BibitemShut {NoStop}%
\bibitem [{\citenamefont {Ping}\ \emph {et~al.}(2014)\citenamefont {Ping},
  \citenamefont {Zhong}, \citenamefont {Sheng} \emph {et~al.}}]{PingPRE2014}%
  \BibitemOpen
  \bibfield  {author} {\bibinfo {author} {\bibfnamefont {Y.~L.}\ \bibnamefont
  {Ping}}, \bibinfo {author} {\bibfnamefont {J.~Y.}\ \bibnamefont {Zhong}},
  \bibinfo {author} {\bibfnamefont {Z.~M.}\ \bibnamefont {Sheng}},  \emph
  {et~al.},\ }\href@noop {} {\bibfield  {journal} {\bibinfo  {journal} {Phys.
  Rev. E}\ }\textbf {\bibinfo {volume} {89}},\ \bibinfo {pages} {031101}
  (\bibinfo {year} {2014})}\BibitemShut {NoStop}%
\bibitem [{\citenamefont {Wang}\ \emph {et~al.}(2000)\citenamefont {Wang},
  \citenamefont {Bhattacharjee},\ and\ \citenamefont {Ma}}]{WangJGR2000}%
  \BibitemOpen
  \bibfield  {author} {\bibinfo {author} {\bibfnamefont {X.}~\bibnamefont
  {Wang}}, \bibinfo {author} {\bibfnamefont {A.}~\bibnamefont {Bhattacharjee}},
  \ and\ \bibinfo {author} {\bibfnamefont {Z.~W.}\ \bibnamefont {Ma}},\ }\href
  {\doibase 10.1029/1999ja000357} {\bibfield  {journal} {\bibinfo  {journal}
  {J. Geophys. Res.}\ }\textbf {\bibinfo {volume} {105}},\ \bibinfo {pages}
  {27633} (\bibinfo {year} {2000})}\BibitemShut {NoStop}%
\bibitem [{\citenamefont {Fox}\ \emph {et~al.}(2017{\natexlab{a}})\citenamefont
  {Fox}, \citenamefont {Sciortino}, \citenamefont {Stechow}, \citenamefont
  {Jara-Almonte}, \citenamefont {Yoo}, \citenamefont {Ji},\ and\ \citenamefont
  {Yamada}}]{FoxPRL2017}%
  \BibitemOpen
  \bibfield  {author} {\bibinfo {author} {\bibfnamefont {W.}~\bibnamefont
  {Fox}}, \bibinfo {author} {\bibfnamefont {F.}~\bibnamefont {Sciortino}},
  \bibinfo {author} {\bibnamefont {Stechow}}, \bibinfo {author} {\bibfnamefont
  {J.}~\bibnamefont {Jara-Almonte}}, \bibinfo {author} {\bibfnamefont
  {J.}~\bibnamefont {Yoo}}, \bibinfo {author} {\bibfnamefont {H.}~\bibnamefont
  {Ji}}, \ and\ \bibinfo {author} {\bibfnamefont {M.}~\bibnamefont {Yamada}},\
  }\href {\doibase 10.1103/physrevlett.118.125002} {\bibfield  {journal}
  {\bibinfo  {journal} {Phys. Rev. Lett.}\ }\textbf {\bibinfo {volume} {118}},\
  \bibinfo {pages} {125002} (\bibinfo {year} {2017}{\natexlab{a}})}\BibitemShut
  {NoStop}%
\bibitem [{\citenamefont {Hesse}\ and\ \citenamefont
  {Winske}(1998)}]{HesseJGP1998}%
  \BibitemOpen
  \bibfield  {author} {\bibinfo {author} {\bibfnamefont {M.}~\bibnamefont
  {Hesse}}\ and\ \bibinfo {author} {\bibfnamefont {D.}~\bibnamefont {Winske}},\
  }\href
  {https://www.scopus.com/inward/record.uri?eid=2-s2.0-0001209926&partnerID=40&md5=b4a62dd8308972093a39f4a7c7b050cd}
  {\bibfield  {journal} {\bibinfo  {journal} {J. Geophys. Res.}\ }\textbf
  {\bibinfo {volume} {103}},\ \bibinfo {pages} {26479} (\bibinfo {year}
  {1998})}\BibitemShut {NoStop}%
\bibitem [{\citenamefont {Rygg}\ \emph {et~al.}(2008)\citenamefont {Rygg},
  \citenamefont {S\'{e}guin}, \citenamefont {Li} \emph
  {et~al.}}]{RyggScience2008}%
  \BibitemOpen
  \bibfield  {author} {\bibinfo {author} {\bibfnamefont {J.~R.}\ \bibnamefont
  {Rygg}}, \bibinfo {author} {\bibfnamefont {F.~H.}\ \bibnamefont
  {S\'{e}guin}}, \bibinfo {author} {\bibfnamefont {C.~K.}\ \bibnamefont {Li}},
  \emph {et~al.},\ }\href {\doibase 10.1126/science.1152640} {\bibfield
  {journal} {\bibinfo  {journal} {Science}\ }\textbf {\bibinfo {volume}
  {319}},\ \bibinfo {pages} {1223} (\bibinfo {year} {2008})}\BibitemShut
  {NoStop}%
\bibitem [{\citenamefont {Li}\ \emph {et~al.}(2010)\citenamefont {Li},
  \citenamefont {S\'{e}guin}, \citenamefont {Frenje} \emph
  {et~al.}}]{LiScience2010}%
  \BibitemOpen
  \bibfield  {author} {\bibinfo {author} {\bibfnamefont {C.~K.}\ \bibnamefont
  {Li}}, \bibinfo {author} {\bibfnamefont {F.~H.}\ \bibnamefont {S\'{e}guin}},
  \bibinfo {author} {\bibfnamefont {J.~A.}\ \bibnamefont {Frenje}},  \emph
  {et~al.},\ }\href {\doibase 10.1126/science.1185747} {\bibfield  {journal}
  {\bibinfo  {journal} {Science}\ }\textbf {\bibinfo {volume} {327}},\ \bibinfo
  {pages} {1231} (\bibinfo {year} {2010})}\BibitemShut {NoStop}%
\bibitem [{\citenamefont {{Retin{\`o}}}\ \emph {et~al.}(2007)\citenamefont
  {{Retin{\`o}}}, \citenamefont {{Sundkvist}}, \citenamefont {{Vaivads}} \emph
  {et~al.}}]{RetinoNP2007}%
  \BibitemOpen
  \bibfield  {author} {\bibinfo {author} {\bibfnamefont {A.}~\bibnamefont
  {{Retin{\`o}}}}, \bibinfo {author} {\bibfnamefont {D.}~\bibnamefont
  {{Sundkvist}}}, \bibinfo {author} {\bibfnamefont {A.}~\bibnamefont
  {{Vaivads}}},  \emph {et~al.},\ }\href {\doibase 10.1038/nphys574} {\bibfield
   {journal} {\bibinfo  {journal} {Nature Physics}\ }\textbf {\bibinfo {volume}
  {3}},\ \bibinfo {pages} {236} (\bibinfo {year} {2007})}\BibitemShut {NoStop}%
\bibitem [{\citenamefont {Opher}\ \emph {et~al.}(2011)\citenamefont {Opher},
  \citenamefont {Drake}, \citenamefont {Swisdak} \emph
  {et~al.}}]{OpherAPJ2011}%
  \BibitemOpen
  \bibfield  {author} {\bibinfo {author} {\bibfnamefont {M.}~\bibnamefont
  {Opher}}, \bibinfo {author} {\bibfnamefont {J.~F.}\ \bibnamefont {Drake}},
  \bibinfo {author} {\bibfnamefont {M.}~\bibnamefont {Swisdak}},  \emph
  {et~al.},\ }\href {http://stacks.iop.org/0004-637X/734/i=1/a=71} {\bibfield
  {journal} {\bibinfo  {journal} {Astrophys. J.}\ }\textbf {\bibinfo {volume}
  {734}},\ \bibinfo {pages} {71} (\bibinfo {year} {2011})}\BibitemShut
  {NoStop}%
\bibitem [{\citenamefont {Matsumoto}\ \emph {et~al.}(2015)\citenamefont
  {Matsumoto}, \citenamefont {Amano}, \citenamefont {Kato},\ and\ \citenamefont
  {Hoshino}}]{MatsumotoScience2015}%
  \BibitemOpen
  \bibfield  {author} {\bibinfo {author} {\bibfnamefont {Y.}~\bibnamefont
  {Matsumoto}}, \bibinfo {author} {\bibfnamefont {T.}~\bibnamefont {Amano}},
  \bibinfo {author} {\bibfnamefont {T.~N.}\ \bibnamefont {Kato}}, \ and\
  \bibinfo {author} {\bibfnamefont {M.}~\bibnamefont {Hoshino}},\ }\href
  {\doibase 10.1126/science.1260168} {\bibfield  {journal} {\bibinfo  {journal}
  {Science}\ }\textbf {\bibinfo {volume} {347}},\ \bibinfo {pages} {974}
  (\bibinfo {year} {2015})}\BibitemShut {NoStop}%
\bibitem [{\citenamefont {Weibel}(1959)}]{WeibelPRL1958}%
  \BibitemOpen
  \bibfield  {author} {\bibinfo {author} {\bibfnamefont {E.~S.}\ \bibnamefont
  {Weibel}},\ }\href {\doibase 10.1103/PhysRevLett.2.83} {\bibfield  {journal}
  {\bibinfo  {journal} {Phys. Rev. Lett.}\ }\textbf {\bibinfo {volume} {2}},\
  \bibinfo {pages} {83} (\bibinfo {year} {1959})}\BibitemShut {NoStop}%
\bibitem [{\citenamefont {Germaschewski}\ \emph {et~al.}(2016)\citenamefont
  {Germaschewski}, \citenamefont {Fox}, \citenamefont {Abbott} \emph
  {et~al.}}]{GermaschewskiJCP2016}%
  \BibitemOpen
  \bibfield  {author} {\bibinfo {author} {\bibfnamefont {K.}~\bibnamefont
  {Germaschewski}}, \bibinfo {author} {\bibfnamefont {W.}~\bibnamefont {Fox}},
  \bibinfo {author} {\bibfnamefont {S.}~\bibnamefont {Abbott}},  \emph
  {et~al.},\ }\href {\doibase 10.1016/j.jcp.2016.05.013} {\bibfield  {journal}
  {\bibinfo  {journal} {J. Comp. Phys.}\ }\textbf {\bibinfo {volume} {318}},\
  \bibinfo {pages} {305} (\bibinfo {year} {2016})}\BibitemShut {NoStop}%
\bibitem [{\citenamefont {Fox}\ \emph {et~al.}(2017{\natexlab{b}})\citenamefont
  {Fox}, \citenamefont {Matteucci}, \citenamefont {Moissard} \emph
  {et~al.}}]{FoxArxiv2017}%
  \BibitemOpen
  \bibfield  {author} {\bibinfo {author} {\bibfnamefont {W.}~\bibnamefont
  {Fox}}, \bibinfo {author} {\bibfnamefont {J.}~\bibnamefont {Matteucci}},
  \bibinfo {author} {\bibfnamefont {C.}~\bibnamefont {Moissard}},  \emph
  {et~al.},\ }\href@noop {} {} (\bibinfo {year} {2017}{\natexlab{b}}),\ \Eprint
  {http://arxiv.org/abs/arXiv:1712.00152} {arXiv:1712.00152} \BibitemShut
  {NoStop}%
\bibitem [{\citenamefont {Hu}\ \emph {et~al.}(2013)\citenamefont {Hu},
  \citenamefont {Michel}, \citenamefont {Edgell} \emph
  {et~al.}}]{SuxingPoP2013}%
  \BibitemOpen
  \bibfield  {author} {\bibinfo {author} {\bibfnamefont {S.}~\bibnamefont
  {Hu}}, \bibinfo {author} {\bibfnamefont {D.~T.}\ \bibnamefont {Michel}},
  \bibinfo {author} {\bibfnamefont {D.~H.}\ \bibnamefont {Edgell}},  \emph
  {et~al.},\ }\href@noop {} {\bibfield  {journal} {\bibinfo  {journal} {Phys.
  Plasmas}\ }\textbf {\bibinfo {volume} {20}},\ \bibinfo {pages} {032704}
  (\bibinfo {year} {2013})}\BibitemShut {NoStop}%
\bibitem [{\citenamefont {Comisso}\ and\ \citenamefont
  {Bhattacharjee}(2016)}]{ComissoJPP2016}%
  \BibitemOpen
  \bibfield  {author} {\bibinfo {author} {\bibfnamefont {L.}~\bibnamefont
  {Comisso}}\ and\ \bibinfo {author} {\bibfnamefont {A.}~\bibnamefont
  {Bhattacharjee}},\ }\href@noop {} {\bibfield  {journal} {\bibinfo  {journal}
  {J. Plasma Phys.}\ }\textbf {\bibinfo {volume} {82}},\ \bibinfo {pages}
  {595820601} (\bibinfo {year} {2016})}\BibitemShut {NoStop}%
\end{thebibliography}%

\end{document}